\documentclass[letterpaper]{JHEP3}
\usepackage{amsmath, amsfonts}
\usepackage{epsfig}
\usepackage{cite}

\def\be{\begin{equation}}
\def\ee{\end{equation}}

\title{Strongly bound mesons at finite temperature and in magnetic fields from AdS/CFT}

\author{K.~D.~Jensen, A.~Karch, J.~Price
\\
Department of Physics, University of Washington,
Seattle, WA 98195-1560 \\
\email{
kristanj@u.washington.edu, karch@phys.washington.edu; cfgauss@u.washington.edu}
}

\abstract{
We study mesons in
${\cal N}=4$ super Yang-Mills theory with fundamental
flavors added at large 't Hooft coupling
using the gauge/gravity correspondence. High-spin mesons
are well described by using semiclassical string configurations.
We determine the meson spectrum
at finite temperature and in a
background magnetic field.
}

\keywords{AdS/CFT correspondence, thermal field theory}

\begin{document}

\section{Introduction}

The determination of the bound state spectrum in a Coulomb
potential is one of the oldest and
most important successes of quantum mechanics. Since the fine-structure
constant in nature is small the binding energy is small,
$E_{bind} \sim \alpha^2 m << m$.
The bound states are essentially non-relativistic. It is interesting to understand what
the properties of bound states look like in the opposite limit of extremely strong Coulomb
interactions.
One system that realizes such strongly coupled Coulombic bound states is ${\cal N}=4$ supersymmetric
Yang-Mills theory with ${\cal N}=2$ supersymmetric flavors in the limit
of large number of colors $N_c$ and large 't Hooft coupling.
The ${\cal N}=4$ theory is conformal,
which guarantees that the potential between external test particles is of the form
$V(r) = -\frac{c_1}{r}$.
In the large $N_c$, large 't Hooft
coupling limit the theory can be solved using the
AdS/CFT correspondence \cite{jthroat,EW,GKP}. In particular the coefficient $c_1$ in
the Coulomb potential between external test sources has been
determined\footnote{We are using conventions where $\lambda =
2 g^2_{YM} N = 4 \pi g_s N$.}
to be
\be
\label{c1}
c_1 = \frac{4 \pi^2}{\Gamma(\frac{1}{4})^4}
\sqrt{\lambda}
\ee
from a Wilson line computation
\cite{mwilson,rwilson}.

To realize Coulombic bound states, one needs to add $N_f$ flavors of
dynamical fundamental
matter (quarks) to the ${\cal N}=4$ theory. In the large $N_c$, finite $N_f$
limit this can be achieved via adding flavor
probe branes on the supergravity side \cite{KarchKatz}. In this
case conformal invariance alone doesn't completely fix the form of the potential.  The quantum corrected Coulomb potential instead has the form
$$V(r) = -\frac{c_1}{r} f(m r).$$
where $m$ denotes the mass of the fundamental flavors.
At strong coupling, the function $f$
can be determined from a static semiclassical string calculation in the bulk.
 This was first done in \cite{myers}. The resulting
potential is depicted in figure (\ref{potential}). At large separation
(alternatively, at large mass) we need to recover the result for infinitely
heavy test quarks and hence $f \rightarrow 1$. At small separation, $f \sim
m^2 r^2$ so that the potential becomes linear in distance, $V(r) = \tau_{eff} r$
with
\be
\label{taueff}
\tau_{eff} = m^2 \frac{2 \pi}{\sqrt{\lambda}}.
\ee
The system is
highly relativistic in this regime: the glue forms flux tubes that are well described
by a relativistic string.

\begin{FIGURE}[t]
    {
    \centerline{\psfig{figure=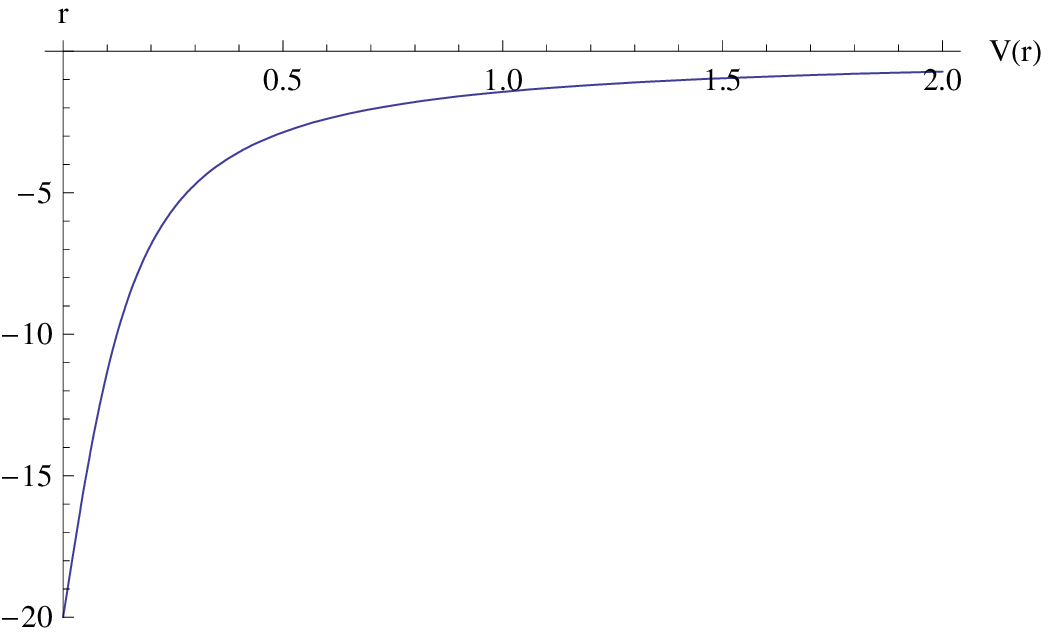,width=3.0in}}
    \caption
    {Quark/antiquark potential $V(r)$
in units where the quark mass is $m=\frac{\sqrt{\lambda}}{2 \pi} 10$.
     }
    \label{potential}
    }
 \end{FIGURE}

At parametrically small separations, $m r \sim
\frac{1}{\sqrt{\lambda}}$ the determination of $f$ in terms of a
semiclassical string breaks down. To ensure that the mesons we study
are insensitive to the details of the potential in this ultra-short
distance regime we study mesons with large angular momentum, $J>>1$.
For $J >> \sqrt{\lambda}$ the quark/antiquark separation is in the
Coulombic regime and the corresponding mesons reproduce
hydrogen-like behavior.  For $1<<J << \sqrt{\lambda}$ the mesons are
dominated by the linear potential and one can compare the spectrum
to that of open relativistic strings in flat space. In all cases the
meson masses are of order the quark mass. On top of this spectrum of
highly spinning strings, there are deeply bound mesons with $J=0,1$
that are dual to the small fluctuations of the flavor brane
worldvolume fields \cite{myers}. Their masses are of order
$\frac{m}{\sqrt{\lambda}}$ and they have been analyzed before at
finite temperature \cite{Babington:2003vm,Albash:2006ew,
Mateos:2007vn,Hoyos:2006gb,Myers:2007we} and non-vanishing magnetic
field \cite{Filev:2007gb, Filev:2007qu,Erdmenger:2007bn} or finite
chemical potential \cite{Erdmenger:2007ja}. For
a review see \cite{Erdmenger:2007cm}. Our studies complement
these results for $J=0,1$ supergravity mesons by analyzing the highly
spinning mesons with semiclassical strings in AdS.

In the next section we start with a review
of the meson properties at zero temperature.
In section 3 we then analyze mesons
and their dissociation at finite temperature. The results are very
similar to those obtained in \cite{Peeters:2006iu} for the
closely related Sakai-Sugimoto model \cite{Sakai:2004cn}.
In section 4 we study mesons
in a background magnetic field and compare to results
for the Zeeman splitting both in hydrogen and for a relativistic
string in the appropriate limits. Finally, in section 5, we bring
the two together and study the
dissociation of mesons at finite temperature in the
presence of a magnetic field.

\section{Meson properties at zero temperature}

Highly spinning mesons at zero temperature were first studied
in \cite{myers}. The corresponding bulk configuration
is a string with both of its ends on the flavor brane. For
large angular momentum $J>>1$, the length of the string
is much larger than $l_s$ and it can be analyzed using
the classical Nambu-Goto action in an AdS background.
Writing the background AdS metric (setting the curvature radius to 1) as
\be
ds^2 = - h(r) dt^2 + r^2 d\vec{x}^2 + \frac{dr^2}{h(r)}
\ee
with $h(r)=r^2$
we are interested in stationary string configurations whose endpoints
describe a circular orbit in the $\vec{x}$ directions. Writing
$ d\vec{x}^2 = d \rho^2 +\rho^2 d \theta^2$ we are hence looking
for a string whose worldvolume is described by
\be
\label{ansatz}
\theta = \omega t, \,\,\,\,\,\,\,\,\,\,
r=r(\sigma), \,\,\,\,\,\,\,\,\,\,
\rho=\rho(\sigma).
\ee
The Nambu-Goto action for this ansatz becomes
\be
S= -\frac{\sqrt{\lambda}}{2 \pi} \int dt d\sigma
\sqrt{\left (r'^2 + \rho'^2 r^2 h(r) \right )
\left ( 1 - \frac{\omega^2 \rho^2 r^2}{h(r)} \right ) } =
 -\frac{\sqrt{\lambda}}{2 \pi} \int dt d\sigma
\sqrt{ (1 - \omega^2 \rho^2)(r'^2 + r^4 \rho'^2)}
\ee
where we used that in the units where the AdS curvature radius is one, the
string tension is simply
\be \frac{1}{2 \pi \alpha'} = \frac{\sqrt{\lambda}}{2 \pi}. \ee
The energy and angular momentum of the string follow immediately
\begin{eqnarray}
J &=& \frac{ \partial L}{\partial \omega} = \frac{\sqrt{\lambda}}{2 \pi}
 \int d\sigma \,  \omega \rho^2 \sqrt{ \frac{r'^2 + r^4 \rho'^2}{1 - \omega^2
\rho^2}} \\
E &=& \omega \frac{ \partial L}{\partial \omega} - L =
 \frac{\sqrt{\lambda}}{2 \pi}
 \int d\sigma \,   \sqrt{ \frac{r'^2 + r^4 \rho'^2}{1 - \omega^2
\rho^2}} .
\end{eqnarray}
The equation of motion is straightforward. The Neumann
boundary condition on $\rho$
that follows from the variation of this Lagrangian is
\be
\label{pi1rho}
 0 = \pi^1_{\rho} = \frac{\partial L}{\partial \rho'} =
\frac{\sqrt{\lambda}} {2 \pi} \frac{\rho ' r^4}{\omega \rho} \sqrt{\frac{1- \omega^2 \rho^2}{r'^2 + r^4
\rho'^2}}
\ee
which simply demands $\rho'=0$ at the end, that is the string
ends on the flavor brane at a right angle at both ends.
Since the configurations
we are interested in are
even under reflection
around the string midpoint,
one can alternatively specify a boundary condition each at the string
midpoint and at a single endpoint.

To proceed, one can further fix the $\sigma$ reparametrization
invariance. Two convenient gauge choices
seem to be either the $\sigma=\rho$ or the $\sigma=r$ static gauge.
Both lead to large gradients which makes the numerics
unstable.
For the numerics we found that a very efficient choice is
\be \sigma = \rho + r \ee
and hence $\rho'+r'=1$. Numerically, we impose boundary conditions
in the IR, demanding that a turnaround point exists at $\rho =0$.
That is, at $\sigma =r_0$ we impose the boundary conditions
\be r(r_0) = r_0, \,\,\,\,\, r'(r_0) = 0 \,\,\,\,
\Rightarrow \,\,\,\,  \rho(r_0) =0, \,\,\,\,\, \rho'(r_0) = 1. \ee
For every choice of turnaround position $r_0$ one then
looks for the first $r_m > r_0$ at which the UV boundary condition
$\rho'(r_m)=0$ is satisfied. This way one avoids shooting and every
numerical run will generate a valid solution for some mass $m$
that is given in terms of $r_m$ by
\be
\label{r_m}
 m =
\frac{\sqrt{\lambda}}{2 \pi} r_m.
\ee
With this numerical procedure we were able to reproduce the
spectrum of mesons $E(J)$ that was previously obtained in
\cite{myers}. Our numerical results, together with the analytic limits
we are about to discuss, are displayed in figure (\ref{myersplot}).
We use the same gauge and numerical methods for our
analysis in later sections.

As discussed in the introduction there are two regimes
in which we expect to be able to compare the meson spectrum
against analytic formulas. For large $J >> \sqrt{\lambda}$
we are in a Coulombic regime where the potential $V(r) = - c_1/r$ with
$c_1$ given in eq.(\ref{c1}). The system is non-relativistic in this regime. The standard non-relativistic
boundstate spectrum
in a Coulomb potential is given by (including the rest-mass $2m$ of the
two quarks)
\be E(J)/m = 2  - \frac{c_1^2}{2 J^2} \frac{m^*}{m}  =
2 - \frac{4 \pi^4}{\Gamma(\frac{1}{4})^8} \frac{\lambda}{J^2}
\ee
where we used that the reduced
mass
is $m^* = \frac{m_1 m_2}{m_1 + m_2} =\frac{m}{2}$ for two equal mass particles.
The other tractable region is $1 << J << \sqrt{\lambda}$ in which
we expect standard Regge behavior
\be E^2 = 2 \pi \tau_{eff} J \ee
where the effective tension is given by eq.(\ref{taueff}) so that
\be E/m = 2 \pi \frac{\sqrt{J}}{\lambda^{1/4}}. \ee

\begin{FIGURE}[t]
    {
    \centerline{\psfig{figure=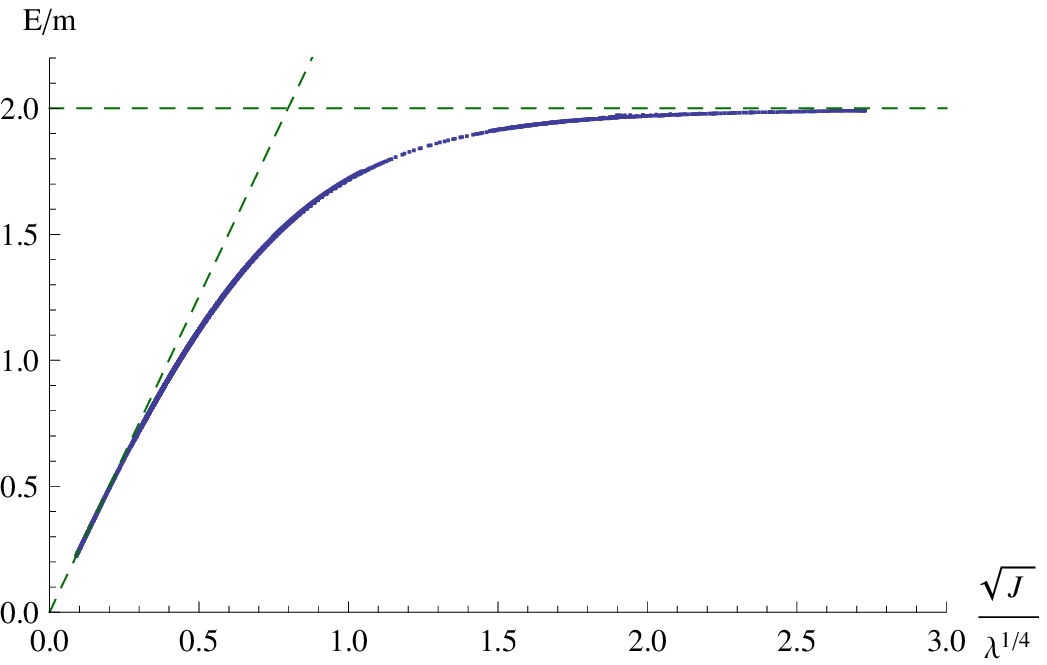,width=3in}}
    \caption
    {Meson energy $E/m$ as a function of
$\sqrt{J}/\lambda^{1/4}$. $E/m=2$ is the rest mass of the quarks, so bound states
only exist for $E/m\leq2$. Also indicated are the two limiting case of
a pure Coulomb potential which is a good approximation for
$J >> \sqrt{\lambda}$ and Regge behavior for $1<<J<<\sqrt{\lambda}$.
     }
    \label{myersplot}
    }
 \end{FIGURE}

\section{Meson dissociation at finite temperature}

In this section we study how the spectrum of highly spinning
mesons is modified
at finite temperature. In particular, we study the dissociation
of the most weakly bound mesons as the temperature is raised.  The disappearance of these bound states leads
to a critical $J_{crit}(m/T)$, where mesons with $J>J_{crit}$ become
unstable. The mesons of flavored ${\cal N}=4$ have
been proposed as a model for charmonium boundstates
at finite temperature e.g. in reference
\cite{Liu:2006nn} where their dissociation as a function
of linear velocity has been analyzed. The study
of meson dissociation is then potentially relevant for studies
at heavy ion colliders. Meson dissociation in the closely
related Sakai-Sugimoto model has been analyzed in
\cite{Peeters:2006iu} with qualitatively similar results.

In the numerical analysis, the only difference with the zero
temperature case is that in the action we need to set $h(r) = r^2-
\frac{r_h^4}{r^2}$ with $r_h = \pi T$ instead of the $h(r)=r^2$ we
had before. In interpreting the result, some care has to be taken in
what one means by mass. As discussed in \cite{Herzog:2006gh} at
finite temperature (and the same comment will later apply to finite
magnetic field) there are several notions of mass which all
coincided in the $B=T=0$ case. One is the mass parameter that
appears in the Lagrangian, $m_L$. This is related to the position
$r_m$ at which the flavor brane ends in a non-trivial fashion
as first discussed in \cite{Babington:2003vm}. One
needs to know the exact embedding of the brane, which changes both
in response to the temperature and the magnetic field. These embeddings
have been obtained in \cite{Erdmenger:2007bn,Albash:2007bq} A physical
definition of mass that is directly related to $r_m$ is the rest
mass of a quasi-particle, which was denoted $M_{rest}$ in
\cite{Herzog:2006gh}. This is simply the energy that one has to pay
to create a flavored quasi-particle and hence just given by the
length of the string stretching from the horizon to the flavor brane
times the tension. For our purposes, we will continue referring to
$m=\frac{\sqrt{\lambda}}{2 \pi} r_m$ as ``the mass''. In the
language of \cite{Herzog:2006gh} this is $M_{rest} + \Delta m$,
latter being the thermal mass shift, $\Delta m =
\frac{\sqrt{\lambda}}{2} T$ . All other notions of mass can simply
be determined in terms of $m$, for details see \cite{Herzog:2006gh}.
With this definition of mass the minimal mass
is $M_{rest}=0$ and hence $m=\Delta m = \frac{\sqrt{\lambda}}{2} T$. This
is the point where the brane
touches the horizon. This minimal value of $m$ appears in
several of our plots.

\begin{FIGURE}[t]
    {
    \centerline{\psfig{figure=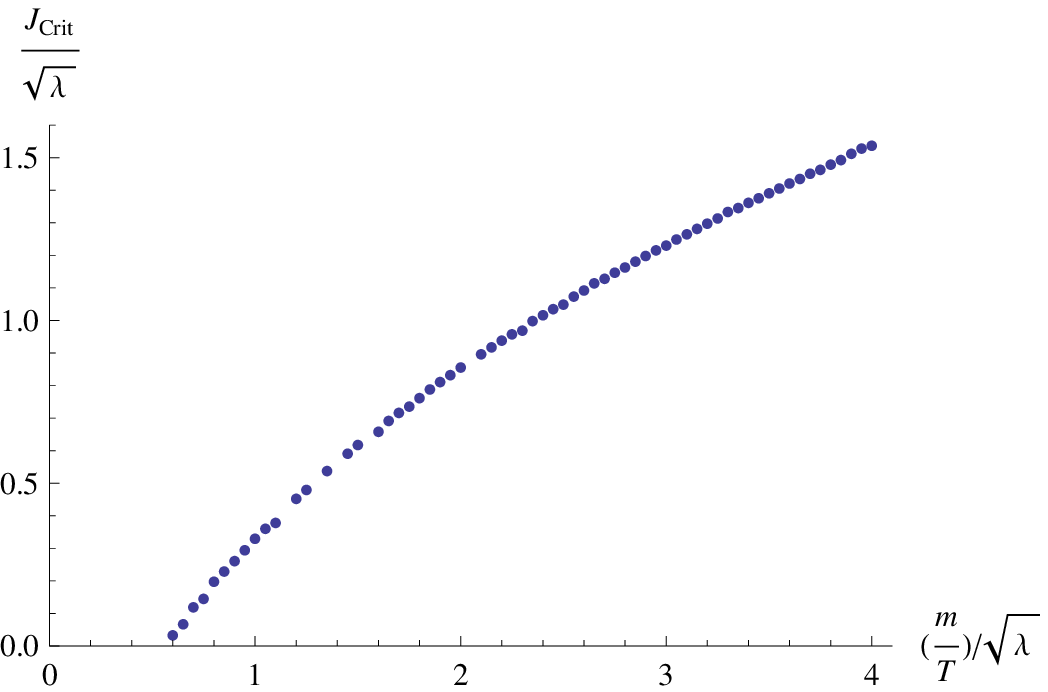,width=3.0in}}
    \caption
    {
$J_{crit}$ as a function of $\frac{m}{T}.$
     }
    \label{jcrit}
    }
 \end{FIGURE}

Figure (\ref{jcrit}) displays our result for the critical angular
momentun as a function of $m/T$.
As in the zero temperature case, one might be tempted to give an
interpretation of $J_{crit}(m/T)$ using a field theory
model of two massive quarks moving in an effective potential. However, such an interpretation is only sensible in the
non-relativistic regime. At finite temperature, due to screening,
the quark/anti-quark potential becomes short range and the
non-relativistic regime gets removed first. For relativistic
bound-states, a significant fraction of the angular momentum and
energy gets carried by the flux tube for which we only have a 5d
description. Qualitatively, the fact that $J_{crit}$ drops as the
temperature rises is consistent with an interpretation of bound
states melting.

Last but not least note that our semiclassical string configurations
only probe the short distance part of the quark/anti-quark potential
that scales as $\sqrt{\lambda}$. As argued in \cite{Bak:2007fk}
there will be an exponential tail on top of this with an order one
coefficient. All we can say with confidence is that for angular
momenta above $J_{crit}$ mesons with binding energies of order
$\sqrt{\lambda}$ do not exist. There might be bound states with
order one binding energies which would be represented by spinning
strings stabilized by quantum corrections.

\section{Zeeman effect}
\subsection{General Considerations}

In this section we study the effect of a background magnetic field
on the meson spectrum. At zero magnetic field rotational symmetry
ensures that the energy of the mesons only depends on the magnitude
$J$ of the angular momentum and not its direction. As usual, the
magnetic field breaks this degeneracy and introduces additional
dependence of the meson masses on the scalar product $\vec{J} \cdot
\vec{B} = m_j B$. With this definition of $m_j$, it takes values
between $-J$ and $+J$ as usual. Since we are working in the
limit of large $J$ we do not see the quantization of $m_j$. For a
generic orientation of $\vec{J}$ and $\vec{B}$, the orbits of the
quark/antiquark pair will no longer be circular and our ansatz for
the bulk string configuration is too simple-minded. However, the
same ansatz eq.(\ref{ansatz}) can still capture the two extreme
cases of $m_j = \pm J$, that is the magnetic field is either
completely aligned or anti-aligned with the angular momentum. These
two extreme cases are sufficient to analyze the splitting in the
magnetic field background. For small $B$, one can linearize the
energies in $\vec{J} \cdot \vec{B} $ and find the standard behavior
\be
\label{standard} \Delta E = a(J) \, m_j B. \ee Calculating the
energy splitting for $m_j = \pm J$ is completely sufficient to
determine $a(J)$. When non-linearities in $B$ become important, we
can still exactly trace the two extreme levels. Values
of $-J<m_j<J$ are expected to lie in between. They need no longer be
equally spaced.

There are actually two cases we will discuss and need to
distinguish: quark/anti-quark pairs with equal opposite charges or
with equal charges. Of course they will always be oppositely charged
under the gauged $SU(N)$ symmetry. For a single flavor brane, there
is just one global $U(1)$ gauge field living on the flavor brane,
baryon number. The quark and the anti-quark carry equal and opposite
charge under this global symmetry. Since quark and anti-quark have
the same mass, the orbital magnetic moment of the two actually
cancels and, like in positronium, the linear Zeeman splitting does
not affect states with different values of $m_j$. To see any
non-trivial linear Zeeman splitting we hence also study the case of
equally charged string endpoints. One way to achieve this is to look
at 2 flavor branes A and B, so that the global flavor symmetry is
$U(1)_A \times U(1)_B$ and then turn on equal opposite fields in the
two U(1) factors. While the AA and BB strings will, like before,
have equal opposite charges under this field, the AB or BA strings
will have equal charges. For simplicity, we only look at 2 flavors
of equal mass, but obviously our method can be extended to the case
of two unequal mass flavors. The corresponding field theory and
gravity configurations are sketched and contrasted in figure
(\ref{bconf}).

\begin{FIGURE}[t]
    {
    \centerline{\psfig{figure=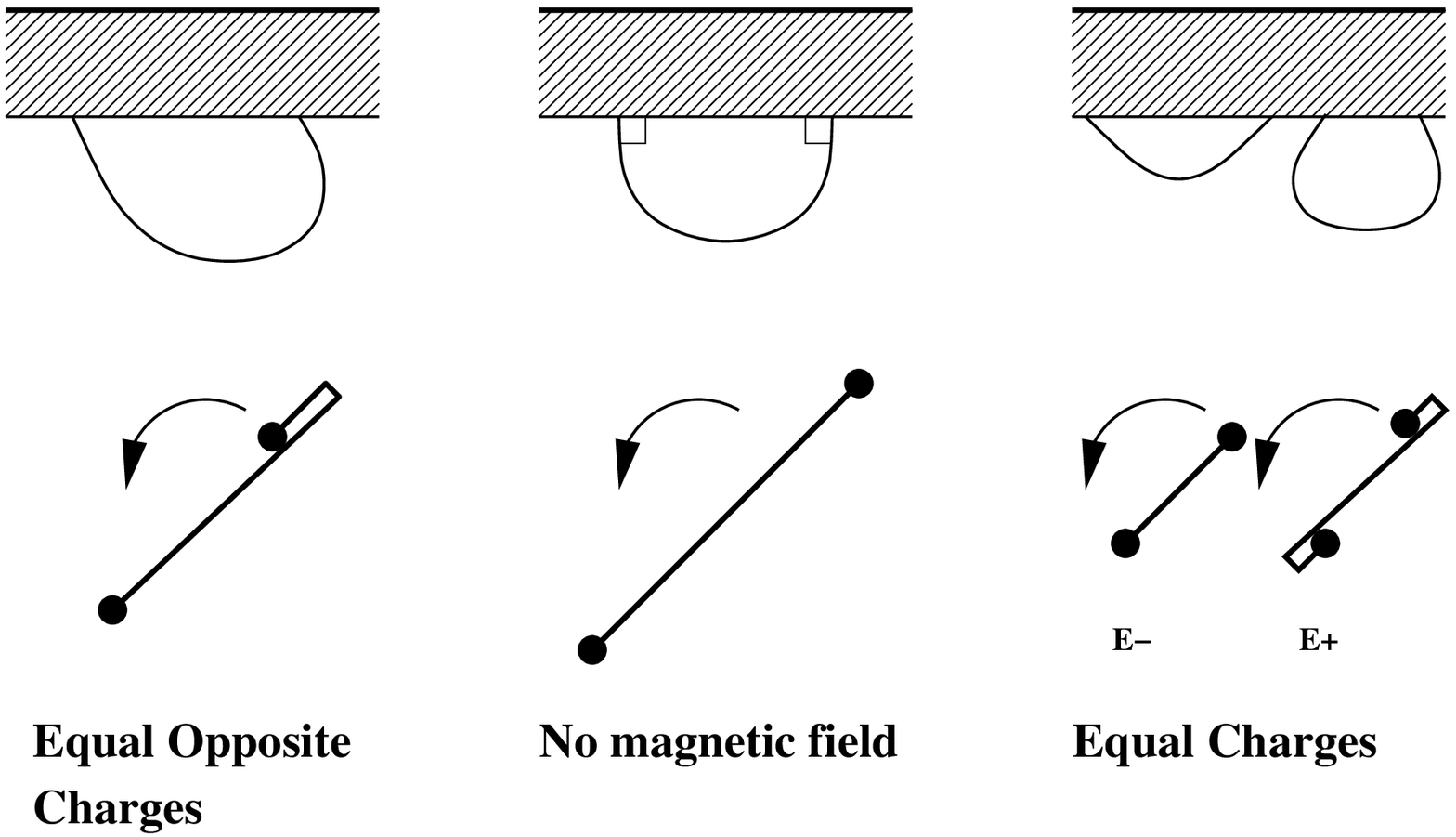,width=4.8in}}
    \caption
    {Sketch of supergravity and field theory configurations
for the cases of equal opposite and equal charges of the string
endpoints. In the field theory picture the circles indicate
the position of quark and anti-quark, whereas the thick lines are the
flux-tubes. In the case of equal opposite charges there is no net
magnetic moment since the contributions from quark and anti-quark
cancel.
In this case
there is a second configuration
with the same energy that is the mirror image of the one depicted.
 In the case of equal charges the two contributions
to the orbital angular momentum
add up. There are two configurations with energies $E_{\pm} =
E \pm
\Delta E$.
     }
    \label{bconf}
    }
 \end{FIGURE}

In the case of equal opposite charges, there is no linear Zeeman splitting
since the magnetic moments of the two particles cancel. There can however be a
correction to the energies quadratic in $B$. Since the only two configurations
we can realize in the supergravity correspond to $\vec{J} \cdot \vec{B} =
\pm JB$, in this case there will be no splitting and the two configurations
will have identical energies. It is indeed easy to see both in the field
theory and the supergravity that in the case of equal opposite charges,
there is a $Z_2$ symmetry that takes $\vec{J} \rightarrow - \vec{J}$
and at the same time exchanges the endpoints of the string. This
symmetry operation takes
one configuration into a different one with identical energy
but opposite sign of $\vec{J} \cdot \vec{B}$. For the case
of $1 << J << \sqrt{\lambda}$ we can compare our supergravity analysis
with the study of a Nambu-Goto string in a background magnetic
field. This was analyzed in the classic paper \cite{Abouelsaood:1986gd}
where it was found that in the case of equal opposite charges
there is no change in the string spectrum whatsoever, even though
the actual string configuration changes: the string starts
to fold back onto itself as indicated in the left of
figure (\ref{bconf}). So for $J <<
\sqrt{\lambda}$ the energy shift as a function of $B^2$ has to vanish.
For $J >> \sqrt{\lambda}$ one expects a non-trivial shift in the
energy, even though there will not be any splitting
between the two configurations with opposite $\vec{J} \cdot \vec{B}$.

In the case of equal charges, we should see the familiar pattern
of Zeeman splitting with a shift proportional to $\vec{J} \cdot \vec{B}$
so that for any given $J$ and $B$ we will find two configurations with circular
orbit and energies $E(B=0) \pm \Delta E$ with $\Delta E$ of the form
indicated in eq. (\ref{standard}). At $J>>\sqrt{\lambda}$ we can compare
our results to an analysis of a purely Coulombic Hydrogen-like system.
In this case we expect
\be \frac{\Delta E}{m} = \frac{\mu_B J B}{m} = \frac{ B}{2 m^2} J \ee
where $\mu_B = \frac{1}{2 m}$ is the standard Bohr magneton for particles
of unit charge.
$\frac{E}{m}$ and $\frac{B}{m^2}$ are the dimensionless quantities
measuring the energy and the magnetic field in units of the quark mass.
Since our theory is conformal, the physics can only depend on these
ratios and not on $E$, $B$ and $m$ individually.
For $1 << J << \sqrt{\lambda}$ one can once more compare to the
spectrum of a Nambu-Goto string with tension $\tau_{eff}$
 in a background magnetic field.
For equal charges it was found in \cite{Abouelsaood:1986gd} that for large $J$
one has
\be E^2 = 2 \pi \tau_{eff} (1 - \epsilon) J \ee
where
\be \epsilon = \frac{2}{\pi} \arctan(\frac{B}{\tau_{eff}}) =
\frac{2}{\pi} \frac{B}{\tau_{eff}} + { \cal O} (B^3) \ee
so that to linear order in $B$ we expect for the change in energy
\be \label{splitting} \frac{\Delta E}{m} = 2 \frac{1}{\sqrt{2 \pi \tau_{eff}}} \sqrt{J}
\frac{B}{m} =
 \frac{B}{ \pi m^2} \lambda^{1/4} \sqrt{J}. \ee

\subsection{Supergravity analysis }

The presence of a constant magnetic field in the $(\rho,\theta)$
plane with $F= B dx \wedge dy =
B \rho dr \wedge d\theta$
only modifies the string action
via the additional boundary terms.
\be
\label{boundaryaction}
\Delta S = \left . \int d\tau A \right |_{\sigma^+} \mp
\left . \int d\tau A \right |_{\sigma^-} \ee
where $\sigma^{\pm}$ refers to the right and left boundary (which
in our gauge choice is not simply at $\sigma = \pi,0$)
and the upper (lower) sign corresponds to the case of equal opposite
(equal) charges.
Choosing $A_{\rho}=0$ gauge we have
as the only non-vanishing component
\be A_{\theta} = B \frac{\rho^2}{2}\ee
and the action (\ref{boundaryaction}) becomes (noting
that $\frac{d \theta}{dt} = \omega$)
\be
\Delta S = \left . B \int d\tau \frac{\rho^2 \omega}{2} \right |_{\sigma^+} \mp
\left . B \int d\tau  \frac{\rho^2 \omega}{2} \right |_{\sigma^-}. \ee
This leads to two basic changes when compared to the analysis
of section 2. First, the boundary condition gets modified to
\be \pi^1_{\rho} = - F_{\rho \theta} \, \dot{\theta} = -
B \rho \omega. \ee
Second, the boundary terms in the action give an additional
contribution to $J = \frac{\partial L}{\partial \omega}$
\be \label{DJ} \Delta J = \left . B \frac{\rho^2}{2} \right |_{\sigma^+} \mp
\left . B \frac{\rho^2 }{2} \right |_{\sigma^-} \ee
The corresponding contribution to the energy
$E = \omega \frac{\partial L}{\partial \omega} - L$ cancels
since the boundary action is linear in $\omega$.
A similar discussion appears in \cite{Arean:2005ar} where mesons
in a non-commuative theory where analyzed which corresponds to a gravity
solution with non-trivial background H-flux.

For the case of two equal charges,
the configurations are once again symmetric under
reflection around the midpoint and one
can then simply perform the numerical analysis along
the same lines as in the zero temperature, zero B-field
case. For every choice of turnaround point $r_0$ one
generates a string configuration that solves the equation of
motion. For every choice of $r_m$ one then can subsequently read
off a $B$ that allows one to have the boundary condition satisfied
at that value of $r_m$. The change in angular momentum in this case is given by $\Delta J = \left . B \rho^2 \right |_{\sigma^+}$.
We present our results for the meson spectrum in this case as
a function of $B/m^2$ together with a comparison to the
analytic results for the $1<<J<<\sqrt{\lambda}$ and $J>>\sqrt{\lambda}$
regimes in figure (\ref{equal}).

\begin{FIGURE}[t]
    {
    \centerline{\psfig{figure=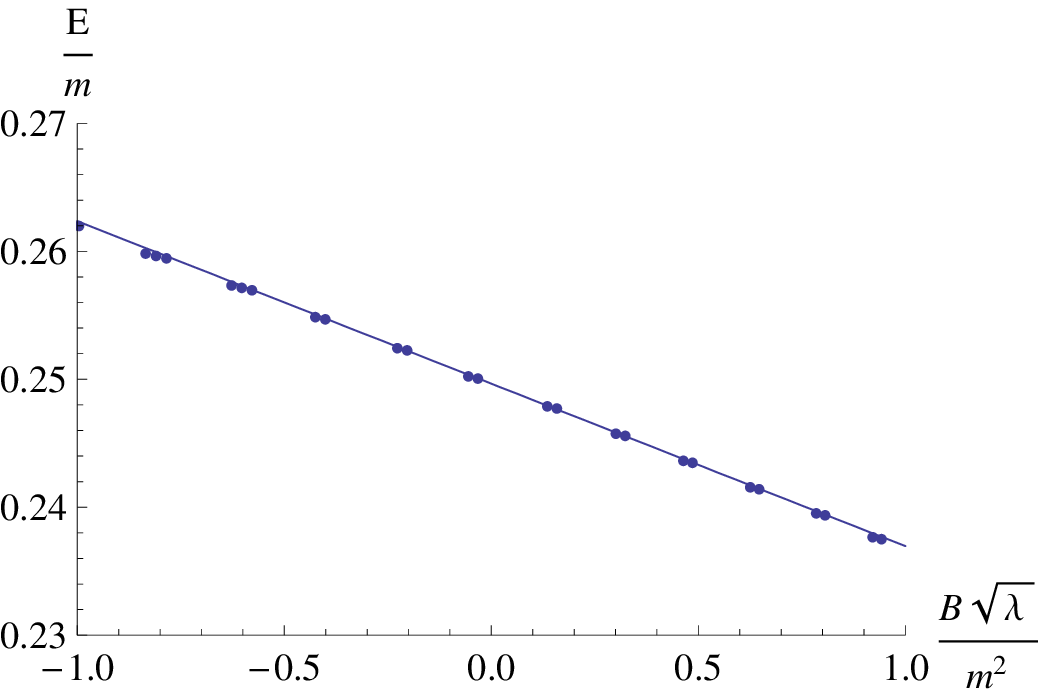, width=2.8in} \psfig{figure=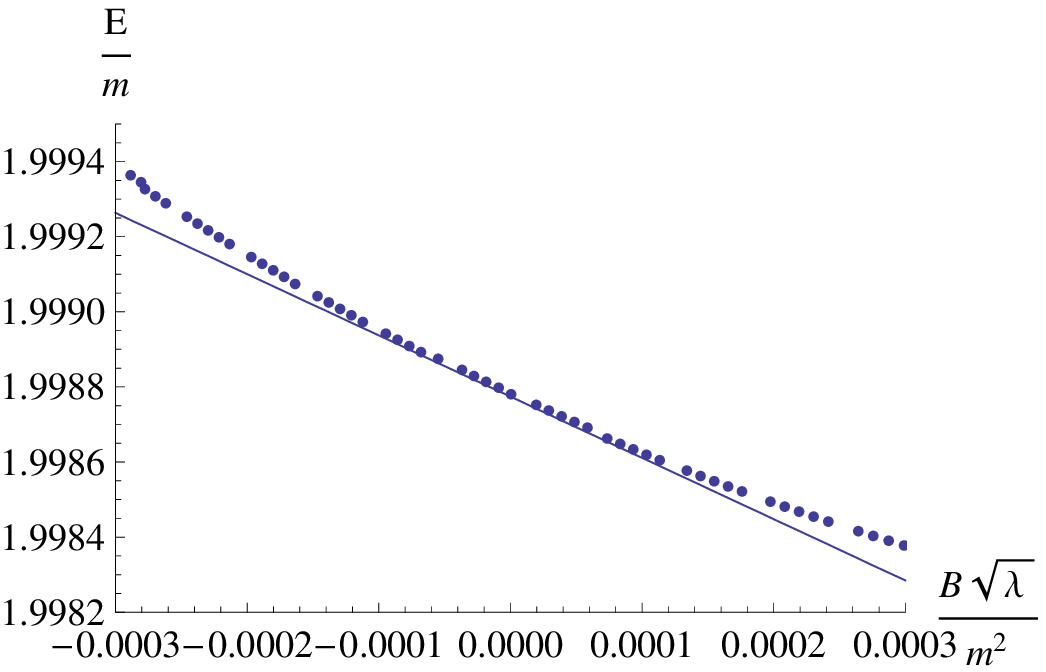, width=2.8in}}
    \caption
    {
Numerical results for
$E/m$ as a function of $B/m^2$
for $J/\sqrt{\lambda} =0.00159$ and for $J/\sqrt{\lambda}=3.34$
together with
analytic
results from the Regge and Coulob regimes respectively in the case
of equal charges.
     }
    \label{equal}
    }
 \end{FIGURE}
In figure (\ref{ofj}) we show the energies for $\frac{B \sqrt{\lambda}}{m^2}= \pm 0.1$
as a function of $J$.
In the left panel of figure (\ref{moredata}) we
also show one example of a calculation performed at an intermediate
$J$ for arbitrary $B$ exhibiting interesting
non-linear effects in $B$.
\begin{FIGURE}[t]
    {
    \centerline{\psfig{figure=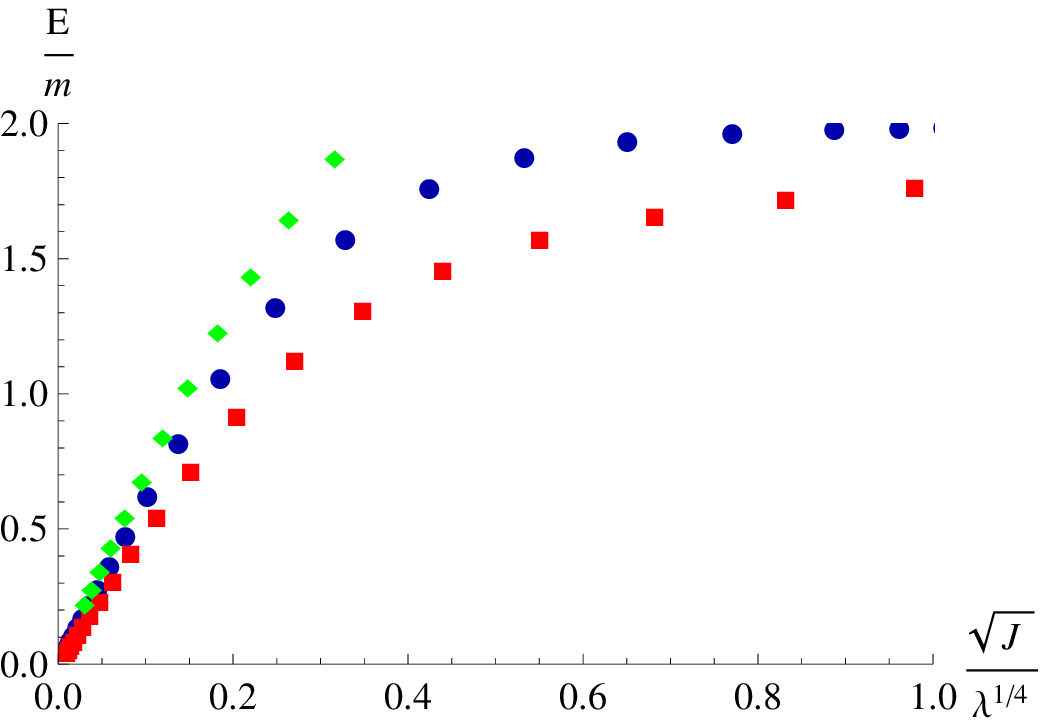, width=3.8in}}
    \caption
    {
$E/m$ as a function of $J$ for $B \sqrt{\lambda}/m^2=0.1$ (blue squares),
$B=0$ (red circles) and $B\sqrt{\lambda}/m^2=-0.1$ (green diamonds).
     }
    \label{ofj}
    }
 \end{FIGURE}

In the case of equal opposite charges one has to deal with the additional
complication that the configuration is no longer symmetric
with respect to reflection around the string midpoint. In
our numerical scheme this means we can no longer assume that the
turnaround point is at $\rho=0$. We therefore must chose both
$r_0$ and $r_0'$ at $\rho=0$. That is at
$\sigma_0= r_0$  we impose the boundary
conditions
\be r(\sigma_0)= r_0, \,\,\,\,\, r'(\sigma_0) =r_0'. \ee
For every $r_m$ one can once more find a $B$ that makes the UV boundary
condition true at either end of the string. But only for
special values of $r_m$ will one get the boundary condition to be
true with the {\it same} $B$ at both ends.
We therefore use a shooting algorithm where, for a given string solution, we step
through possible values of the mass to find the special $r_m$'s for which
the asymmetric boundary conditions are satisfied.
We present our results for the meson spectrum in this case
in the right panel of
figure (\ref{moredata}). A typical string
configuration for this case of equal opposite charges
is displayed in figure (\ref{stringprofile})

\begin{FIGURE}[t]
    {
    \centerline{\psfig{figure=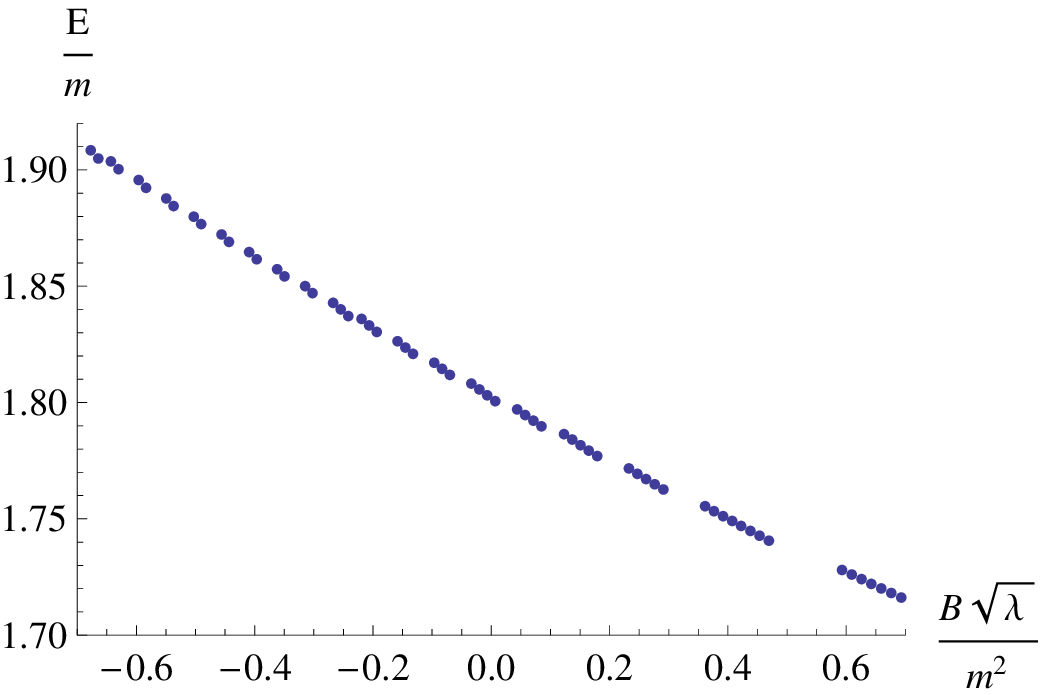, width=2.8in} \psfig{figure=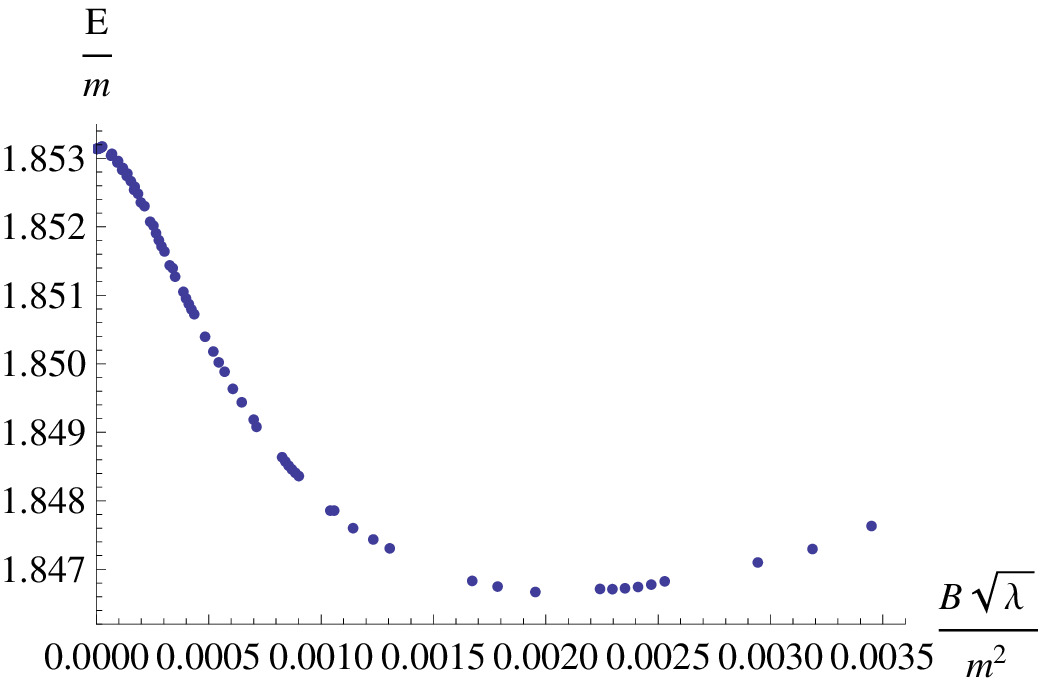, width=2.8in}}
    \caption
    {Left panel: Meson spectrum $E/m$ as a function of $B/m^2$ for $J/\sqrt{\lambda}=0.207$
for equal charges. Right Panel: Meson spectrum $E/m$ as a function
of $B/m^2$ for $J/\sqrt{\lambda}=0.255$ for equal opposite charges.
Note that for equal opposite charges the linear Zeeman effect does
not contribute since the meson has no net magnetic moment, the shift
starts at quadratic order in $B$.
     }
    \label{moredata}
    }
 \end{FIGURE}

\begin{FIGURE}[t]
    {
    \centerline{\psfig{figure=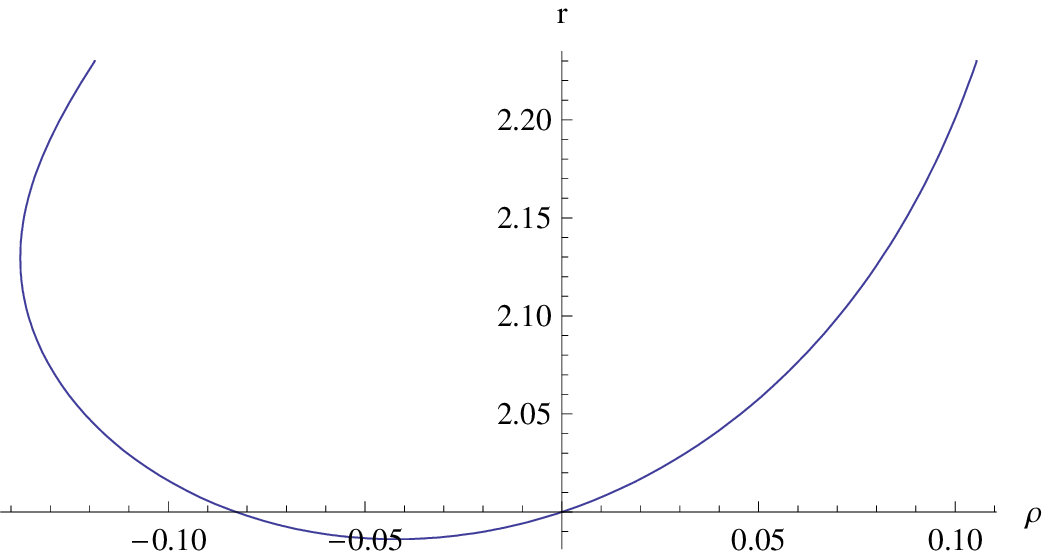,width=3.8in}}
    \caption
{String profile $r(\rho)$ for the case of equal
opposite charges with $\frac{J}{\sqrt{\lambda}}=0.126$ and
$\frac{B}{m^2}=\frac{3.64}{\sqrt{\lambda}}$.
     }
    \label{stringprofile}
    }
\end{FIGURE}

\subsection{$J << \sqrt{\lambda}$, equal charges}

An analytic solution to the equations of motion can be found in the
case $\omega \rightarrow \infty$, following a procedure similar to the one in \cite{myers}.
We will see that this case corresponds to $J << \sqrt{\lambda}$.
To simplify the equations, we use the coordinates
$\tilde{z} = \omega z = \omega/r$, $\tilde{\rho} = \omega \rho$
and the gauge $\tilde{\rho} = \sigma$.  The equation of motion for $\tilde{z}$
is then
\be
\label{EoM}
\frac{\tilde{z}''}{1+\tilde{z}'^{\,2}} + \frac{2}{\tilde{z}} -
\frac{\tilde{\rho}\tilde{z}'}{1-\tilde{\rho}^{\,2}} = 0
\ee
We can substitute the ansatz $\tilde{z} = \omega z_{D7} + \frac{1}{\omega z_{D7}}
f(\tilde{\rho})$ into (\ref{EoM}), where $z_{D7}$ is the position of the D7 brane.
Keeping only terms of order $1/\omega z_{D7}$ we get an equation for $f$,
\be
f'' - 2 - \frac{\tilde{\rho}f'}{1-\tilde{\rho}^2} = 0
\ee
{}From the boundary conditions for $\tilde{z}$, we find the boundary
conditions $f(0)=0$ and $f'(0)=0$.  Then, solving for f,
\be
f(\tilde{\rho}) = \frac{1}{2}(\tilde{\rho}^2 + \arcsin^2(\tilde{\rho}) ).
\ee
At the boundary $\rho$ must satisfy $\pi^1_\rho
= \frac{\partial L}{\partial \rho'}
= - \rho B \omega$.  So,
\be
\frac{\sqrt{\lambda}}{2 \pi}\frac{1}{\tilde{z}^2}\sqrt{ \frac{1-\tilde{\rho}^{\,2}}{1+\tilde{z}'^{\,2}}}
= - \tilde{\rho} B
\ee
Substituting the equation for $\tilde{z}$, we see that the
critical value of $\tilde{\rho}$ at the boundary must satisfy, for small $B$,
\be
\label{rho_C}
\tilde{\rho}_{C} = \frac{\sqrt{\lambda}/{2\pi}}{\sqrt{\lambda/{4 \pi^2}+B^2 z_{D7}^4}} \approx 1
- \frac{B^2 z_{D7}^4 \pi^2 }{\lambda}.
\ee
Putting all of this into the equations for $E$ and $J$, including
the endpoint contribution to $J$ (\ref{DJ}), using that $z_{D7}=\frac{\sqrt{\lambda}}{2 m \pi}$,
and keeping terms linear in $B$, we find that
\be
E = \frac{ \sqrt{\lambda}}{2 z_{D7}^2 \omega}-\frac{2B}{\omega}
\ee
\be
J = \frac{ \sqrt{\lambda}}{4 z_{D7}^2 \omega^2}-\frac{B}{\omega^2}
\ee
Since $\omega$ is large, $J<<\sqrt{\lambda}$, as we expected.
Solving for $E$ in terms of $J$, using (\ref{r_m}) and again keeping only terms
linear in $B$,
\be
\frac{E}{m} = \frac{2 \pi \sqrt{J}}{\lambda^{1/4}} - \frac{\sqrt{J} \lambda^{1/4}}{\pi} \frac{B}{m^2}.
\ee
To find the energy of the upper branch, we must integrate from $\rho_{C}$ to $\rho=1$, and
then from $\rho=1$ back to $\rho=0$ to take into account the shape of the string configurations
as we see in figure (\ref{bconf}).  We find the energy of this branch to be,
\be
\frac{E}{m} = \frac{2 \pi \sqrt{J}}{\lambda^{1/4}} + \frac{\sqrt{J} \lambda^{1/4}}{\pi} \frac{B}{m^2}.
\ee
So the splitting is $\Delta E/m = \sqrt{J} \lambda^{1/4} B/m^2 \pi$,
which is consistant with what was obtained in (\ref{splitting}).

\subsection{$J<<\sqrt{\lambda}$, opposite charges}
For the case of equal and opposite charges, the boundary condition is different for the left and right
ends of the string.
\be
\left. \pi_1^\rho \right|_{\pm} = \pm\rho B \omega
\ee
As we can see from the string profile in figure (\ref{bconf}), on one side of the string, we must again integrate
from $\rho_C$ to $\rho=1$ to $\rho=0$, as in the upper branch of the equal charge case.
The other side of the string is integrated as in the lower branch of the equal charges case.
Following the same method as before, we obtain,
\be
E = \frac{\pi}{z_{D7}^2 \omega}
\ee
\be
J = \frac{\pi}{2 z_{D7}^2 \omega^2}
\ee
so
\be
E/m = \frac{2\pi \sqrt{J}}{\lambda^{1/4}}
\ee
which does not depend linearly on $B$.  
In this case, exact evaluation of the integrals of the small $J$ solution without
using the additional approximation (\ref{rho_C}) shows that $E$ does not depend on $B$. 
To see what this means, 
consider writing $E^2$ as a power series in $J$ with functional prefactors $C_i(B)$,
\be (E/m)^2 = \sum_i C_i(B)J^i. \ee
The result above is simply the statement that $C_1$ is the constant $4\pi^2/\sqrt{\lambda}$. This is in perfect agreement with the
result of \cite{Abouelsaood:1986gd} that for a field theory
Nambu-Goto flux tube with equal
opposite charges the energy is completely independent of $B$.

\section{Meson dissociation in the presence of magnetic fields}

Last but not least, we want to put together the two cases
we studied in the last two sections and study the strings
when both temperature and external magnetic field are turned on.
There are of course now many different parameters to tune: the
temperature, the mass, the magnetic field, and the angular momentum.
The main point we want to make here is that one can map
out the properties of the strings as a function of all those control
parameters using our technique. One particularly interesting quantity
we calculate is once more the critical angular momentum $J_{crit} (m/T,B/m^2)$ beyond
which the mesons dissociate. In figure (\ref{jcritb}) we show $J_{crit}$
as a function of $m/T$  for $B= \pm 0.157 \sqrt{\lambda} T^2$.
As expected,
the magnetic field increases the binding energy of at least one state and hence
increases $J_{crit}$ when compared to the zero temperature cases.

\begin{FIGURE}[t]
    {
   \centerline{\psfig{figure=
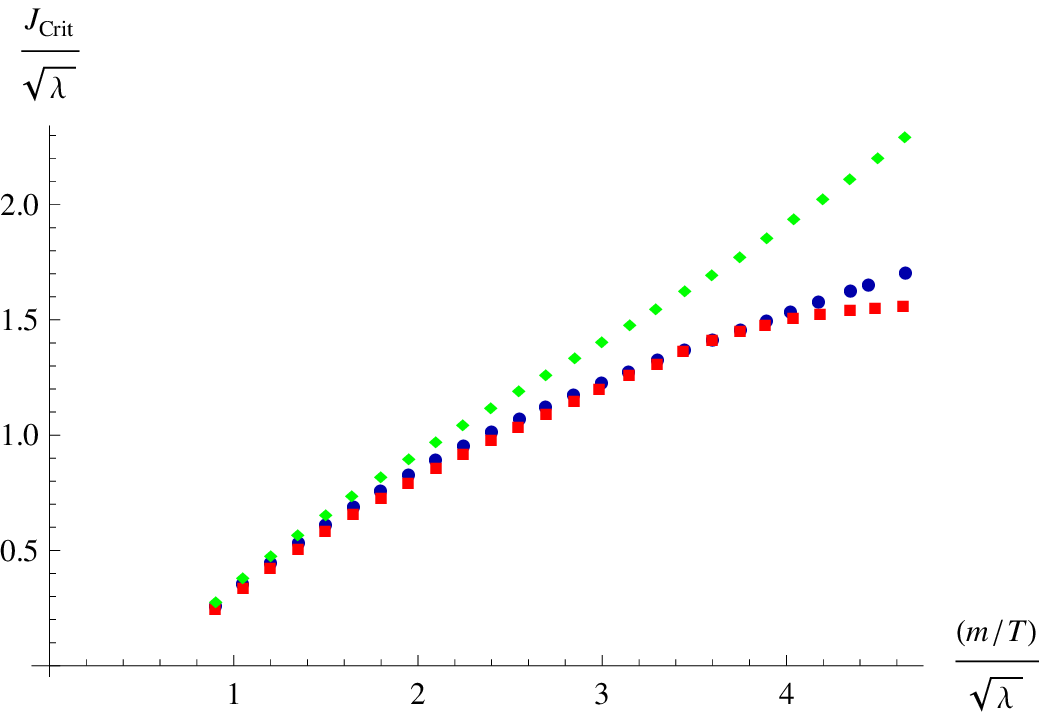,width=3.8in}
}
    \caption
    {
$J_{crit}$ as a function of $m/T$ for $\frac{B}{\sqrt{\lambda}T^2}=-0.157$
 (green diamonds), 0 (blue circles)
and +0.157 (red squares).
     }
    \label{jcritb}
    }
 \end{FIGURE}

\section*{Acknowledgments}
We would like to thank M.~Baker and in particular C.~Herzog for useful discussions.
This work was supported in part by the U.S. Department
    of Energy under Grant No.~DE-FG02-96ER40956.

\bibliography{gkp}

\providecommand{\href}[2]{#2}\begingroup\raggedright\begin{thebibliography}{10}

\bibitem{jthroat}
J.~M. Maldacena, {\it The large {$N$} limit of superconformal field theories
  and supergravity},  {\em Adv. Theor. Math. Phys.} {\bf 2} (1998) 231--252,
  [\href{http://xxx.lanl.gov/abs/hep-th/9711200}{{\tt hep-th/9711200}}].

\bibitem{EW}
E.~Witten, {\it Anti-de {S}itter space and holography},  {\em Adv. Theor. Math.
  Phys.} {\bf 2} (1998) 253--291,
  [\href{http://xxx.lanl.gov/abs/hep-th/9802150}{{\tt hep-th/9802150}}].

\bibitem{GKP}
S.~S. Gubser, I.~R. Klebanov, and A.~M. Polyakov, {\it Gauge theory correlators
  from non-critical string theory},  {\em Phys. Lett.} {\bf B428} (1998)
  105--114, [\href{http://xxx.lanl.gov/abs/hep-th/9802109}{{\tt
  hep-th/9802109}}].

\bibitem{mwilson}
J.~M. Maldacena, {\it Wilson loops in large {$N$} field theories},  {\em Phys.
  Rev. Lett.} {\bf 80} (1998) 4859--4862,
  [\href{http://xxx.lanl.gov/abs/hep-th/9803002}{{\tt hep-th/9803002}}].

\bibitem{rwilson}
S.-J. Rey and J.-T. Yee, {\it Macroscopic strings as heavy quarks in large
  {$N$} gauge theory and anti-de {S}itter supergravity},  {\em Eur. Phys. J.}
  {\bf C22} (2001) 379--394,
  [\href{http://xxx.lanl.gov/abs/hep-th/9803001}{{\tt hep-th/9803001}}].

\bibitem{KarchKatz}
A.~Karch and E.~Katz, {\it Adding flavor to {AdS/CFT}},  {\em JHEP} {\bf 06}
  (2002) 043, [\href{http://xxx.lanl.gov/abs/hep-th/0205236}{{\tt
  hep-th/0205236}}].

\bibitem{myers}
M.~Kruczenski, D.~Mateos, R.~C. Myers, and D.~J. Winters, {\it Meson
  spectroscopy in {AdS/CFT} with flavour},  {\em JHEP} {\bf 07} (2003) 049,
  [\href{http://xxx.lanl.gov/abs/hep-th/0304032}{{\tt hep-th/0304032}}].

\bibitem{Babington:2003vm}
J.~Babington, J.~Erdmenger, N.~J. Evans, Z.~Guralnik, and I.~Kirsch, {\it
  Chiral symmetry breaking and pions in non-supersymmetric gauge/gravity
  duals},  {\em Phys. Rev.} {\bf D69} (2004) 066007,
  [\href{http://xxx.lanl.gov/abs/hep-th/0306018}{{\tt hep-th/0306018}}].

\bibitem{Albash:2006ew}
T.~Albash, V.~Filev, C.~V. Johnson, and A.~Kundu, {\it A topology-changing
  phase transition and the dynamics of flavour},
  \href{http://xxx.lanl.gov/abs/hep-th/0605088}{{\tt hep-th/0605088}}.

\bibitem{Mateos:2007vn}
D.~Mateos, R.~C. Myers, and R.~M. Thomson, {\it Thermodynamics of the brane},
  {\em JHEP} {\bf 05} (2007) 067,
  [\href{http://xxx.lanl.gov/abs/hep-th/0701132}{{\tt hep-th/0701132}}].

\bibitem{Hoyos:2006gb}
C.~Hoyos, K.~Landsteiner, and S.~Montero, {\it Holographic meson melting},
  {\em JHEP} {\bf 04} (2007) 031,
  [\href{http://xxx.lanl.gov/abs/hep-th/0612169}{{\tt hep-th/0612169}}].

\bibitem{Myers:2007we}
R.~C. Myers, A.~O. Starinets, and R.~M. Thomson, {\it Holographic spectral
  functions and diffusion constants for fundamental matter},
  \href{http://xxx.lanl.gov/abs/arXiv:0706.0162 [hep-th]}{{\tt arXiv:0706.0162
  [hep-th]}}.

\bibitem{Filev:2007gb}
V.~G. Filev, C.~V. Johnson, R.~C. Rashkov, and K.~S. Viswanathan, {\it
  Flavoured large n gauge theory in an external magnetic field},
  \href{http://xxx.lanl.gov/abs/hep-th/0701001}{{\tt hep-th/0701001}}.

\bibitem{Filev:2007qu}
V.~G. Filev, {\it Criticality, scaling and chiral symmetry breaking in external
  magnetic field},  \href{http://xxx.lanl.gov/abs/arXiv:0706.3811
  [hep-th]}{{\tt arXiv:0706.3811 [hep-th]}}.

\bibitem{Erdmenger:2007bn}
J.~Erdmenger, R.~Meyer, and J.~P. Shock, {\it Ads/cft with flavour in electric
  and magnetic kalb-ramond fields},  {\em JHEP} {\bf 12} (2007) 091,
  [\href{http://xxx.lanl.gov/abs/arXiv:0709.1551 [hep-th]}{{\tt arXiv:0709.1551
  [hep-th]}}].

\bibitem{Erdmenger:2007ja}
J.~Erdmenger, M.~Kaminski, and F.~Rust, {\it Holographic vector mesons from
  spectral functions at finite baryon or isospin density},
  \href{http://xxx.lanl.gov/abs/arXiv:0710.0334 [hep-th]}{{\tt arXiv:0710.0334
  [hep-th]}}.

\bibitem{Erdmenger:2007cm}
J.~Erdmenger, N.~Evans, I.~Kirsch, and E.~Threlfall, {\it Mesons in
  gauge/gravity duals - a review},
  \href{http://xxx.lanl.gov/abs/arXiv:0711.4467 [hep-th]}{{\tt arXiv:0711.4467
  [hep-th]}}.

\bibitem{Peeters:2006iu}
K.~Peeters, J.~Sonnenschein, and M.~Zamaklar, {\it Holographic melting and
  related properties of mesons in a quark gluon plasma},  {\em Phys. Rev.} {\bf
  D74} (2006) 106008, [\href{http://xxx.lanl.gov/abs/hep-th/0606195}{{\tt
  hep-th/0606195}}].

\bibitem{Sakai:2004cn}
T.~Sakai and S.~Sugimoto, {\it Low energy hadron physics in holographic {QCD}},
   {\em Prog. Theor. Phys.} {\bf 113} (2005) 843--882,
  [\href{http://xxx.lanl.gov/abs/hep-th/0412141}{{\tt hep-th/0412141}}].

\bibitem{Liu:2006nn}
H.~Liu, K.~Rajagopal, and U.~A. Wiedemann, {\it An {AdS/CFT} calculation of
  screening in a hot wind},  {\em Phys. Rev. Lett.} {\bf 98} (2007) 182301,
  [\href{http://xxx.lanl.gov/abs/hep-ph/0607062}{{\tt hep-ph/0607062}}].

\bibitem{Herzog:2006gh}
C.~P. Herzog, A.~Karch, P.~Kovtun, C.~Kozcaz, and L.~G. Yaffe, {\it Energy loss
  of a heavy quark moving through n = 4 supersymmetric yang-mills plasma},
  {\em JHEP} {\bf 07} (2006) 013,
  [\href{http://xxx.lanl.gov/abs/hep-th/0605158}{{\tt hep-th/0605158}}].

\bibitem{Albash:2007bq}
T.~Albash, V.~G. Filev, C.~V. Johnson, and A.~Kundu, {\it {Quarks in an
  External Electric Field in Finite Temperature Large N Gauge Theory}},
  \href{http://xxx.lanl.gov/abs/arXiv:0709.1554 [hep-th]}{{\tt arXiv:0709.1554
  [hep-th]}}.

\bibitem{Bak:2007fk}
D.~Bak, A.~Karch, and L.~G. Yaffe, {\it Debye screening in strongly coupled n=4
  supersymmetric yang-mills plasma},  {\em JHEP} {\bf 08} (2007) 049,
  [\href{http://xxx.lanl.gov/abs/arXiv:0705.0994 [hep-th]}{{\tt arXiv:0705.0994
  [hep-th]}}].

\bibitem{Abouelsaood:1986gd}
A.~Abouelsaood, J.~Callan, Curtis~G., C.~R. Nappi, and S.~A. Yost, {\it Open
  strings in background gauge fields},  {\em Nucl. Phys.} {\bf B280} (1987)
  599.

\bibitem{Arean:2005ar}
D.~Arean, A.~Paredes, and A.~V. Ramallo, {\it Adding flavor to the gravity dual
  of non-commutative gauge theories},  {\em JHEP} {\bf 08} (2005) 017,
  [\href{http://xxx.lanl.gov/abs/hep-th/0505181}{{\tt hep-th/0505181}}].

\end{thebibliography}\endgroup
\bibliographystyle{JHEP}

\end{document}